\documentstyle[twocolumn,prl,aps,psfig]{revtex}
\begin{document}
\title{Phase separation in supersolids}
\author{G. G. Batrouni$^{1}$ and R. T. Scalettar$^2$}
\address{$^1$
Institut Non-Lin\'eaire de Nice,
Universit\'e de Nice--Sophia Antipolis, 
1361 route des Lucioles,
06560 Valbonne, France}
\address{$^2$
Physics Department,
University of California,
Davis CA 95616, USA}
\date{\today}
\address{\mbox{ }}        
\address{\parbox{14cm}{\rm \mbox{ }\mbox{ }     
We study quantum phase transitions in the ground state of the two
dimensional hard--core boson Hubbard Hamiltonian.  Recent work on this
and related models has suggested ``supersolid'' phases with
simultaneous diagonal and off--diagonal long range order.  We show
numerically that, contrary to the generally held belief, the most
commonly discussed ``checkerboard'' supersolid is thermodynamically
unstable.  Furthermore, this supersolid cannot be stabilized by next
near neighbour interaction.  We obtain the correct phase diagram using
the Maxwell construction.  We demonstrate the ``striped'' supersolid
is thermodynamically stable and is separated from the superfluid phase
by a continuous phase transition.}}
\address{\mbox{ }}
\address{\parbox{14cm}{\rm \mbox{ }\mbox{ }
}}
\address{\mbox{ }}
\maketitle
\narrowtext  
\vskip 0.7cm

Nearly forty years ago, Penrose and Onsager~\cite{Penrose} posed the
question: Is it possible for a bosonic system, like $^4$He, to have a
phase where long range crystal order (``solid'') and off diagonal long
range order (superfluidity) co--exist?  Their answer, based on an
analysis which ignored zero--point fluctuations, was that such
supersolid phases do not occur.  Subsequently, it was
argued~\cite{andreev,chester,leggett} that including the effect of
large zero--point quantum fluctuations in the crystal phase might
allow for the existence of a supersolid.  This question has continued
to spark much theoretical, numerical and experimental
interest.\cite{review} There has been a convergence of agreement,
based on mean--field\cite{chester,liufisher,matsuda,ggbsuperPRB} and
numerical\cite{ggbsuperPRB,loh,vanotterlo,ggbsuperPRL,monien} work,
that supersolids do exist, in $2d$ lattice models, particularly in
systems in which the density of bosons is doped away from the
commensurate fillings which are optimal for charge ordering.  The
existence of supersolids for 2-d quantum bosons has fundamental
implications to vortex phases in superconductors because of formal
mappings between the problems.\cite{vortex} In this paper, we
demonstrate that the most discussed of these lattice supersolid phases
is thermodynamically unstable, and argue that it does not exist in any
region of interaction strength or density.

Consider a 2--d square lattice with one hard-core boson for every two
sites ($\rho=\frac12$) interacting with near neighbor (nn)
repulsion. If the interactions are weak, the bosons will be mobile and
condense into a superfluid phase at low temperature.  If repulsion is
strong, the system will freeze into a charge density wave pattern in
which sites are alternately occupied and empty.  At $\rho=\frac12$
these possibilities are mutually exclusive.  If we remove or add a
boson, the resulting bosonic defect could ``hop'' among the background
of charge ordered particles if the zero--point fluctuations are large
enough.  A dilute gas of such defects may Bose condense and form a
superfluid superimposed on the background of crystal order, a
``supersolid'' phase.  If, instead, next near--neighbor (nnn)
repulsion dominates, the charge ordering is in stripes, but the basic
issue of a condensation of additional bosons coexisting with a striped
pattern is as for checkerboard. While it is useful to think of
separate frozen and superfluid bosons, these quantum particles are
indistinguishable.  All the bosons simultaneously participate in both
types of long range order.

Calculations supporting this intuitive physical picture are primarily
based on mean field theory with spin wave stability analysis.  They
initially dealt with checkerboard charge order where the ordering
vector for the structure factor is ${\vec k}=(\pi,\pi)$. Liu and
Fisher~\cite{liufisher} argued that the supersolid phase exists for
hard--core bosons with nn repulsion, but that it is unstable in the
sense that the {\it critical velocity} vanishes.  If nnn repulsion is
present, the supersolid can be stabilized.\cite{RMP} Numerical
simulations of the quantum phase model (QPM),\cite{JJA} which
describes soft-core bosons, showed that the $(\pi,\pi)$ supersolid
phase exists even without nnn repulsion, due to the soft cores.  The
supersolid is present even at half--filling, {\it i.e.~}in the absence
of any defects~\cite{vanotterlo}.  Simulations of the {\it hard--core}
bosonic Hubbard model similarly found that the supersolid phase exists
in the absence of nnn repulsion off half--filling, but unlike the QPM
is absent at half--filling.  In addition, a mean field with spin wave
analysis showed that this supersolid phase has a finite critical
velocity.\cite{ggbsuperPRL,ggbsuperPRB} However, it is generally
accepted that at least in the presence of nnn repulsion, the
$(\pi,\pi)$ supersolid phase is stable.

Discussions of stability based on nonvanishing critical velocity
examine the effects of low energy excitations on an {\it existing}
supersolid phase. To our knowledge, however, there has been no
discussion or numerical verification of the underlying thermodynamic
stability of either of these supersolid phases against phase
separation.  Simulations done at {\it fixed} particle number which
found simultaneous diagonal and off--diagonal long range
order,\cite{ggbsuperPRB,vanotterlo,ggbsuperPRL,roddickstroud,kohno},
do not address the possibility of phase separation.  In what follows
we examine thermodynamic stability of the checkerboard and striped
supersolids by constructing the chemical potential--particle number
relation and calculating the compressibility.

We use a new Dual Quantum Monte Carlo Algorithm\cite{dual} to simulate
the hard--core bosonic Hubbard model, 
\begin{eqnarray}
H=&-&t\sum_{\langle
  ij\rangle}(a_{i}^{\dagger}a_{j}+a_{j}^{\dagger}a_{i})
\nonumber \\
&+&V_{1}\sum_{\langle ij \rangle}\hat
  n_{i}\hat n_{j}+ V_{2}\sum_{\langle\langle ik\rangle\rangle}\hat
  n_{i}\hat n_{k}.
\label{hub-ham}
\end{eqnarray}
$a_i$ ($a_i^{\dagger}$) are destruction (creation) operators of
hard--core bosons on site $i$ of a 2--d square lattice, and $n_i$ is
the density at site $i$.  The hopping parameter is chosen to be $t=1$
to fix the energy scale.  $V_1$ ($V_2$) is the near neighbor (next
near neighbor) interaction.  At $V_2=0$, and after an appropriate
sublattice spin rotation, this boson model is equivalent to the
spin--$\frac12$ antiferromagnetic XXZ model.  In this language,
superfluid order corresponds to magnetic order in the XY plane, while
density order corresponds to magnetic order in the Z direction.

\begin{figure}
\hspace*{-10mm}
\psfig{file=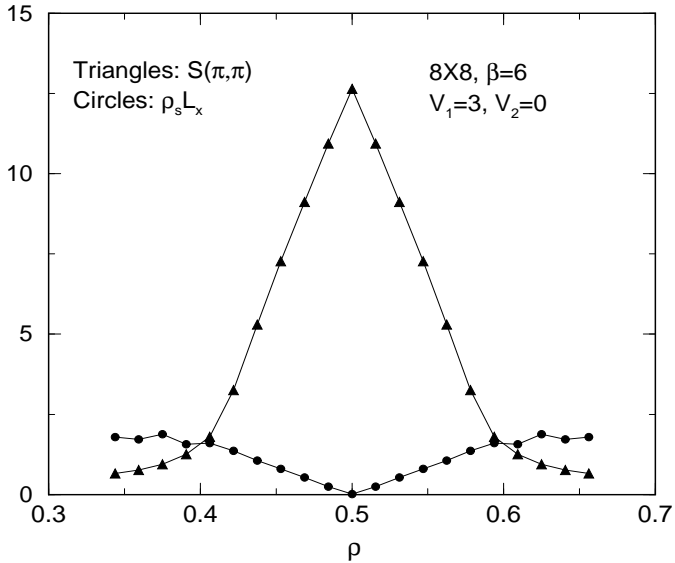,height=3.0in,width=3.5in,angle=-90}
\vskip-00mm
\caption{ 
The structure factor, $S(\pi,\pi)$, and $\rho_s$ as a function of
fixed density.  The half-filled point $\rho=0.5$ is a solid with
$\rho_s=0$.  For $\rho$ close to $\rho=0.5$, $S(\pi,\pi)$ and $\rho_s$
are both nonzero. Moving away from half-filling, eventually
$S(\pi,\pi)$ will no longer scale linearly with system size and the
system is a pure superfluid.  }
\end{figure}         

Traditionally, to determine numerically the nature of the ground state
of (1), we evaluate, at fixed density, the superfluid density,
$\rho_s$, and the equal time structure factor at the ordering vector
${\bf q}$, 
\begin{equation} S({\bf q}) = {1 \over N} \sum_{{\bf l}}
e^{i {\bf q} \cdot {\bf l}} \langle n({\bf j},\tau)n({\bf j}+{\bf
l},\tau)\rangle.  
\end{equation} 
Ground state results for $S({\bf q})$ and $\rho_s$ are shown in Fig.~1
(Fig.~2) for the checkerboard (striped) phase.  In both cases $\rho_s$
is nonzero everywhere except precisely at half--filling, but $S({\bf
q})$ also remains large off half--filling, indicating solid order.
Using finite size scaling to extrapolate to the limit of infinite
lattice size for these fixed density
systems,\cite{ggbsuperPRB,vanotterlo,ggbsuperPRL,roddickstroud,kohno}
one can show that density correlations are indeed still long ranged
off half-filling where $\rho_s \neq 0$.  The conclusion is that
checkerboard and striped supersolid phases exist in the thermodynamic
limit.

Already, however, it was remarked~\cite{ggbsuperPRB} that the energy
versus density curves had small negative curvature in the ($\pi,\pi$)
supersolid phase. It was speculated that this was evidence for phase
separation. The $(\pi,0)$ supersolid phase showed no such negative
curvature. It was recently shown numerically~\cite{kohno} that the
easy--axis spin-$1/2$ XXZ model on a square lattice exhibits a first
order spin--flop transition. These results confirm the discontinuous
nature of the transition from the superfluid phase as $\rho$ is
adjusted in the {\it absence} of nnn repulsion.

\begin{figure}
\hspace*{-10mm}
\psfig{file=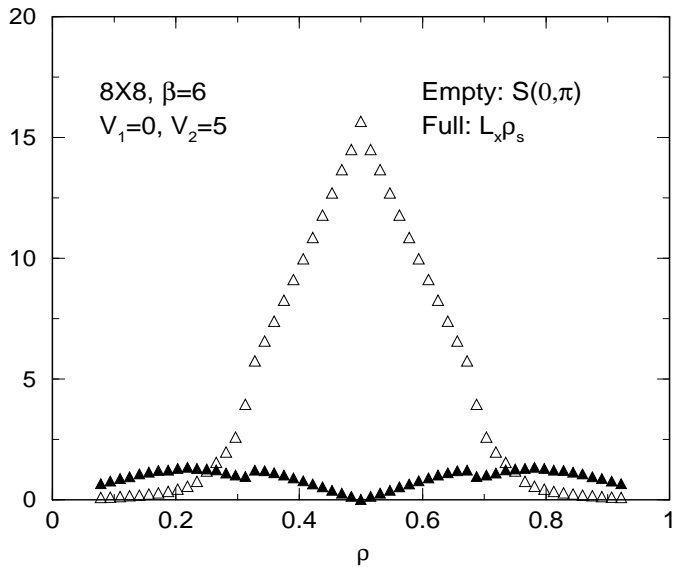,height=3.0in,width=3.5in,angle=-90}
\vskip-00mm
\caption{ 
The structure factor $S(0,\pi)$ and $\rho_s$ as a function of fixed
density.  The half-filled point $\rho=0.5$ is a solid with $\rho_s=0$.
For $\rho$ close to $\rho=0.5$, $S(0,\pi)$ and $\rho_s$ are both
nonzero.  As with the checkerboard case, finite size scaling for
$S(0,\pi)$ determines the density at which solid order vanishes.
}
\end{figure}         

To address the possibility of phase separation systematically, {\it
i.e.}~to obtain the phase diagram in the interaction--chemical
potential ($\mu$) plane, we must obtain $\rho$ as a function of $\mu$,
and use the Maxwell construction~\cite{max}.  We first study the
checkerboard case by fixing $V_2=0$ and scanning the filling for
several values of $V_1$. The chemical potential for $n$ bosons is
calculated from the total energy: $\mu(n)=E(n+1)-E(n)$.  We work on
lattices up to size 12x12, and temperatures as low as $\beta=6$ to
access the ground state properties.

Fig.~3 shows $\rho$ versus $\mu$ for $V_1=3, V_2=0$, as in Fig.~1. The
slope of this curve is the compressibility, $\kappa={\partial \rho /
\partial \mu}$. Two $\kappa< 0$ branches are clearly seen just before
and after the energy gap. The gap itself corresponds to the
incompressible $(\pi,\pi)$ solid at half filling and seen in
Fig.~1. Using the Maxwell construction we find the critical value of
the chemical potential, $\mu_c$ (vertical dashed line), and read off
the critical filling $\rho_c$ in Fig.~3.  The structure factor
(Fig.~1) begins a very rapid rise at the point where $\kappa$ turns
negative.\cite{ggbtocome} It is crucial to note that $\rho_s$ and
$S(\pi,\pi)$ are both non-zero, Fig.~1, {\it only} in the {\it
unstable} $\kappa< 0$ region (Fig.~3).  On one side of the transition
the phase is purely superfluid while on the other side it is a
($\pi,\pi$) solid.  What was previously accepted to be the
checkerboard supersolid at fixed density, lies entirely on the
$\kappa<0$ branches and therefore phase separates into a mixture of
solid and superfluid phases at densities $\rho=\rho_c$ and
$\rho=\frac12$. The metastable states in Fig.~3 correspond either to
the superfluid or the gapped insulating phases.

\begin{figure}
\hspace*{-10mm}
 \psfig{file=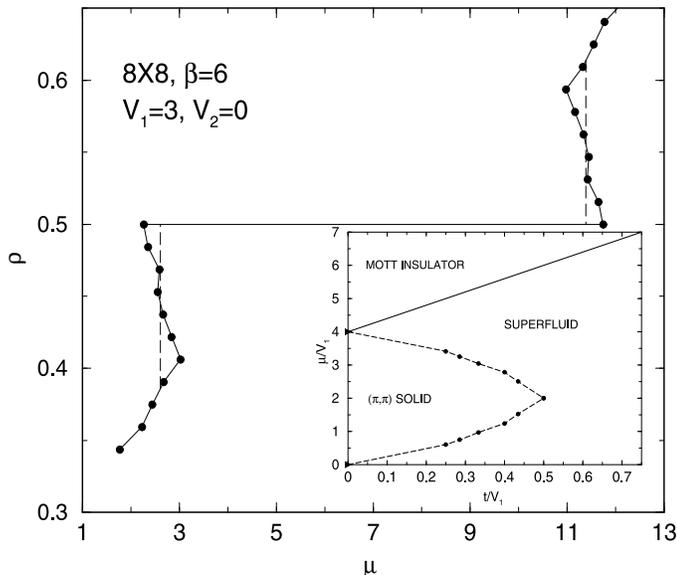,height=3.0in,width=3.5in,angle=-90}
\vskip-00mm
\caption{
$\rho$ versus $\mu$ showing the $\kappa < 0$
regions. The vertical dashed line shows the location of the
transitions from the Maxwell construction.  Inset: The phase diagram
for $V_2=0$. The solid line shows the continuous transition to the
Mott phase at full filling, the dashed line shows the discontinuous
transitions from the superfluid to the checkerboard solid at half
filling. The density changes discontinuously across this line. The tip
of the lobe is a continuous critical point.
}
\end{figure}

To check if nnn repulsion stabilizes this phase against phase
separation, we did simulations~\cite{ggbtocome} with $V_1=3$ and $V_2$
ranging from deep in the $(\pi,\pi)$ solid region to close to the
boundary with striped order. We found the same $\kappa < 0$ behavior
as in Fig.~3.  Next near neighbour repulsion {\it does not} stabilize
the checkerboard supersolid phase.

Repeating the simulations that gave Fig.~3 for different values of
$V_1$ we construct the phase diagram in the ($\mu/V_1, t/V_1, V_2=0$)
plane, shown as the inset to Fig.~3.  As the tip of the lobe is
approached, the energy gap opens without $\kappa < 0$ regions in the
$\rho, \mu$ plane.  Therefore this point is apparently a continuous
transition.  This is consistent with Ref.~\cite{FB} while
Ref.~\cite{RS} finds a first order transition in a model with longer
range (Coulomb) interactions.

The same analysis for $V_1=0$, scanning $V_2$ and $\rho$, determines
the stability of the striped supersolid. Fig.~4 is a
typical plot of $\rho$ versus $\mu$ traversing the incompressible
(gapped) striped solid at $\rho=\frac12$. This is strikingly different
 from Fig.~3. There is no $\kappa < 0$ region: The phase transitions
are all continuous.  Furthermore, as $\mu$ is increased from the
lowest shown value, the slope, {\it i.e.}  $\kappa$, changes markedly
at $(\mu \approx 0.74, \rho \approx 0.25)$.  We find that $S(0,\pi)$
(Fig.~2) begins a rapid increase, indicating long range striped order,
at precisely this particle density, $\rho$.  Since the superfluid
density, $\rho_s$, is still finite, we conclude that the rapid
crossover in $\kappa$, like the behavior of the structure factor,
signals the continuous transition from the superfluid to the ($0,\pi$)
supersolid. Increasing $\mu$ further takes the system into the gapped
($0,\pi$) insulator, then back to the striped supersolid and finally
to the superfluid phase at $\mu \approx 19.2$.  At strong coupling,
the gap (the jump in $\mu$) across the incompressible striped phase is
$4V_2 = 20$ for the 2-d square lattice.  This value is reduced by
quantum fluctuations as $V_2$ decreases, eventually disappearing
entirely at weak coupling.  The absence of negative compressibility
regions indicates that all phases, in particular the striped
supersolid, are thermodynamically stable.

\begin{figure}
\hspace*{-10mm}
 \psfig{file=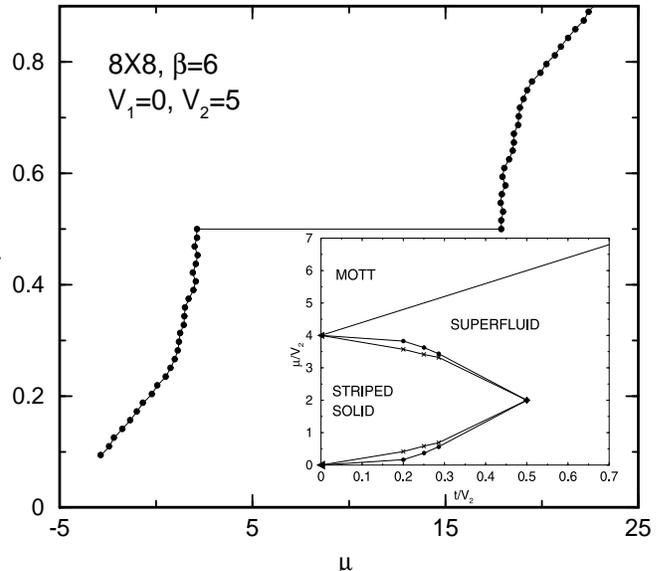,height=3.0in,width=3.5in,angle=-90}
\vskip-00mm
\caption{
$\rho$ versus $\mu$.  Inset: The phase diagram for $V_1=0$. The narrow
regions sandwiched between SF and $(\pi,0)$ solid phases are the
stable supersolid phases.  }
\end{figure}         

Repeating these simulations for various values of $V_2$ gives us the
phase diagram in the ($\mu/V_2, t/V_2, V_1=0$) plane 
(inset to Fig.~4).  The regions sandwiched between the superfluid and
the striped solid are the two stable striped supersolid phases. As the
tip is approached, the supersolid phase gets narrower since the
SS-$(0,\pi)$-solid transition approaches the SS-SF transition. This
prevents us from resolving these transition points near the tip. The
difficulty in the numerical determination of these points stems from
the fact that when the SF--SS and SS-striped solid transitions get
very close to each other, they start behaving numerically as
multicritical points.  However, it appears that the supersolid phase
completely surrounds the striped solid phase except at the tip and the
base where we have multicritical points.

We also studied the effect of nn repulsion on the ($0,\pi$) phase and
found that the $(0,\pi)$ supersolid remains stable and that additional
gapped phases appear at other special
fillings.\cite{vanotterlo,longrange} Whether there are associated
supersolid phases is under investigation~\cite{longrange}.

The fact that it is easier to support nonzero $\rho_s$ in a striped
solid than a checkerboard one can be qualitatively argued as follows:
In a striped solid doped away from half--filling, defects have
channels in which they can move at no interaction energy cost, and,
importantly, the kinetic energy of these defects is set by $t$ and can
be controlled independently of the strength of the interaction $V_2$
which determines the solid order. In a checkerboard solid, the motion
of a defect proceeds through an intermediate state of energy $2V_1$,
giving a reduced effective hopping $t_{{\rm eff}} \approx
t^2/2V_1$. $V_1$ controls simultaneously the defect kinetic energy and
the tendency to charge order.  As a consequence, there is reduced
ability to tune to a supersolid phase. It is still remarkable, though,
that the striped phase forms even at very low densities.
Indeed, in the fermion Hubbard model, very small doping (just
a few percent) away from half-filling destroys long range spin order 
(antiferromagnetism), leaving little possibility that it might
coexist with superconductivity.

In this paper, we have presented Quantum Monte Carlo results for
ground state correlations in the hard--core bosonic Hubbard model with
near-- and next near--neighbor repulsion.  We show that the ${\bf
q}=(\pi,\pi)$ checkerboard supersolid, contrary to current beliefs, is
an unstable phase and does not exist thermodynamically for this model
for any filling or nnn repulsion. Instead, the system phase separates
into solid and superfluid phases. This contradicts mean--field
predictions which examine stability via an evaluation of the critical
velocity for spin waves, and find that in the presence of next near
neighbour repulsion the supersolid is stable.  We have not examined
the soft--core case in detail, but preliminary results indicate
negative compressibility regions in that case too.\cite{ggbsuperPRB}
The quantum phase model for soft core bosons also exhibits negative
compressibility phases.\cite{rieger}

We found the striped supersolid phase, ${\bf q}=(\pi, 0$) to be stable
and separated from the superfluid phase by a continuous transition.
The energy $E(n)$ provides a signal of the transition: The
compressibility exhibits a rapid cross--over from the superfluid to
the supersolid phase, with $\kappa_{SF}< \kappa_{SS}$.  The issue of
the stability of possible supersolid phases at other densities and
wavevectors which are associated with the presence of long range
interactions is a fascinating one which is presently under
investigation.

\centerline{Acknowledgments}
We acknowledge useful discussions with G.~Zimanyi, H.~Rieger, and
M.~Loaf.

\end{document}